\documentclass{article}
\usepackage{arxiv}

\usepackage[utf8]{inputenc} 
\usepackage[T1]{fontenc}    
\usepackage[hidelinks]{hyperref}       
\usepackage{url}            
\usepackage{booktabs}       
\usepackage{amsfonts}       
\usepackage{nicefrac}       
\usepackage{microtype}      
\usepackage{lipsum}		
\usepackage{graphicx}
\usepackage[square,numbers]{natbib}
\usepackage{doi}
\usepackage{caption}
\usepackage{subcaption}
\usepackage{amsmath}

\usepackage{bm}

\usepackage{url}
\usepackage[usenames,dvipsnames]{xcolor}
\usepackage{xcolor}
\definecolor{newcolor}{rgb}{.8,.349,.1}

   \newcommand{\mean}[1]{\overline{#1}\,}
   \newcommand{\pmean}[1]{\overline{\overline{#1}}\,}
   \newcommand{\logmean}[1]{\overline{#1}^{\text{log}}}
   \newcommand{\gmean}[1]{\overline{#1}^{G}}
   \newcommand{\hmean}[1]{\overline{#1}^{H}}

   \newcommand{\tildemean}[1]{\widetilde{#1}\,}

\newtheorem{remark}{Remark}

  \newcommand{\meantwo}[1]{\overline{#1}^{\,\#}}
  \newcommand{\meanone}[1]{\overline{#1}^{\,\star}}

   \newcommand{\mC}{\mathcal{C}}
   \newcommand{\mM}{\mathcal{M}}
   \newcommand{\mP}{\mathcal{P}}
   \newcommand{\mF}{\mathcal{F}}

   \newcommand{\massflux}{\mF_{\rho}}
   
   \usepackage{caption}
   \usepackage{subcaption}
   \usepackage{epsfig,epstopdf}
   \usepackage{ulem}
   
\title{Novel Pressure-Equilibrium and Kinetic-Energy Preserving fluxes for compressible flows based on the harmonic mean}%

\author{ \href{https://orcid.org/0000-0002-6518-3114}{\includegraphics[scale=0.06]{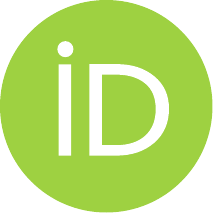}\hspace{1mm} Carlo {De~Michele}}\\
	Dipartimento di Ingegneria Industriale\\
	Universit\`a di Napoli ``Federico II''\\
	Napoli, Italy \\
	\texttt{carlo.demichele2@unina.it} \\
	\And
	\href{https://orcid.org/0000-0003-4943-9551}{\includegraphics[scale=0.06]{orcid.pdf}\hspace{1mm}Gennaro Coppola} \\
	Dipartimento di Ingegneria Industriale\\
	Universit\`a di Napoli ``Federico II''\\
	Napoli, Italy \\
	\texttt{gcoppola@unina.it} \\
}

\renewcommand{\shorttitle}{Novel PEP and KEP fluxes for compressible flows based on the harmonic mean}%

\hypersetup{
pdftitle={Novel Pressure-Equilibrium and Kinetic-Energy Preserving fluxes for compressible flows based on the harmonic mean},
pdfsubject={physics.flu-dyn},
pdfauthor={Carlo De~Michele, Gennaro Coppola}
pdfkeywords={Compressible flows, Turbulent flows, Kinetic-energy preserving methods, Pressure equilibrium preserving method},
}

\begin{document}
\maketitle

\begin{abstract}
Employing physically-consistent numerical methods is an important step  towards attaining robust and accurate numerical simulations.
When addressing compressible flows, in addition to preserving kinetic energy at a discrete level, as done in the incompressible case, additional properties are sought after, such as the ability to preserve the equilibrium of pressure that can be found at contact interfaces.
This paper investigates the general conditions of the spatial numerical discretizations to achieve the pressure equilibrium preserving property (PEP).
Schemes from the literature are analyzed in this respect, and procedures to impart the PEP property to existing discretizations are proposed.
Additionally, new PEP numerical schemes are introduced through minor modifications of classical ones.
Numerical tests confirmed the theory hereby presented and showed that the modifications, beyond the enforcement of the PEP property, have a generally positive impact on the performances of the original schemes.
\end{abstract}

\section{Introduction} \label{sec:Introduction}
Despite the long-standing research efforts starting in the 1950s with the pioneering work at Los Alamos National Laboratory, the design of numerical methods for the discretization of Navier-Stokes equations is still an active research topic. The growth of computational resources continually pushes the community towards more challenging applications, and the design of the `optimal' discretization setting for the various flow regimes, especially for turbulent flows, is still an ambitious goal.

In the context of compressible flows, considerable research efforts 
have focused over many decades on the problem of the correct and efficient representation of admissible weak solutions (i.e.~shock waves or contact discontinuities). However, there are some numerical issues associated with smooth regions of the flow, such as the nonlinear instability 
due to the spatial discretization of convective terms, which have a great impact in numerical simulations at high Reynolds numbers and have received less attention. Only in recent years this topic has seen a renowned interest in the wider context of structure-preserving or physics-compatible methods. These methods aim to devise stable and reliable numerical discretizations without the explicit addition of numerical dissipation, but discretely reproducing the (conservative) structure of suitably selected induced quantities.
The first paper addressing these topics is that of \citet{Feiereisen_1981} in which the authors, inspired by previous studies for incompressible formulations, 
use a skew-symmetric like splitting of the convective terms in mass and momentum equations within the framework of a 
Finite Difference (FD) discretization. This formulation was shown to achieve the so-called Kinetic Energy Preserving (KEP) property, i.e.~the induced discrete equation for kinetic energy has convective terms automatically coming in conservative form, which implies that the discrete formulation does not spuriously produce kinetic energy, with a remarkable increase in reliability of the numerical simulations.
After this seminal work, several other attempts have been made to obtain more general KEP schemes, among which we recall here the Finite Volume (FV) analyses of \citet{Jameson_JSC_2008} and \citet{Subbareddy_JCP_2009} and the works on the triple splitting by \citet{Kennedy_JCP_2008} and \citet{Pirozzoli_JCP_2010}. In more recent years, a quite complete characterization of the possible KEP formulations in a FD framework has been derived~\cite{Coppola_JCP_2019} and its relations with FV formulations based on algebraic fluxes have been explored~\cite{Coppola_JCP_2023,Coppola_ECCOMAS_2022}. 
Among other less standard extensions of interest, we mention the approach of \citet{Edoh_JCP_2022}, utilizing the rotational form of the momentum equation, and that in which a formulation of the governing equations based on square-root variables is used \cite{Morinishi_JCP_2010,Rozema_JT_2014,Nordstrom_JCP_2022,DeMichele_AIAAAviation_2023}.

In addition to kinetic energy, also discrete entropy conservation has been explored as a means to improve fidelity and suppress instabilities in turbulent simulations. 
The first attempt in this direction from a FD standpoint was put forward by \citet{Honein_JCP_2004}, who proposed to integrate a modified (non conservative) version of the total-energy equation corresponding, for exact time integration, to a conservative discretization of the entropy equation.
Although this approach achieves exact entropy conservation at the cost of spoiling the total-energy balance, it has shown increased robustness and a more faithful representation of thermodynamic fluctuations in numerical simulations of isotropic turbulence. 
In the context of FV methods, the theory of entropy variables developed by Tadmor~\cite{Tadmor_MC_1987,Tadmor_AN_2003}, forms the basis for the construction of Entropy Conservative (EC) numerical fluxes that can discretely reproduce the correct entropy balance without sacrificing the total-energy conservation. 
\citet{Ismail_JCP_2009} developed `affordable' EC numerical fluxes based on the logarithmic mean for ideal gases, whereas \citet{Chandrashekar_CCP_2013} proposed for the first time both KEP and EC numerical fluxes as a basis for entropy-stable discretization of Euler and Navier-Stokes equations.
More recently, Ranocha~\cite{Ranocha_2020,Ranocha_CAMC_2021} developed a more refined KEP and EC formulation which is also able to enforce the so-called Pressure Equilibrium Preserving (PEP) property~\cite{Shima_JCP_2021}, i.e.~the ability of the discrete scheme to reproduce the traveling density wave solutions of the compressible Euler equations, obtained for initially uniform pressure and velocity distributions.
This last scheme appears as the most complete one in terms of structure-preserving properties for compressible Euler equations and ideal gases.
However, as it is based on logarithmic means, it comes with a non-negligible increase in computational cost and is potentially singular for uniform distributions of density or temperature. 
These disadvantages have been recently solved by \citet{DeMichele_JCP_2023}, who proposed a family of Asymptotically Entropy Conservative (AEC) schemes which are PEP and KEP and arbitrarily reduce the error on entropy conservation. As they are based on arithmetic and harmonic means, they only use algebraic operations for the computation of the fluxes, with increased efficiency at comparable performances.
Moreover, the fluxes do not exhibit any singularity for uniform distributions.
This family of schemes improves existing formulations which approximately enforce the entropy-conservation property, as the KEEP or KEEP$^{(N)}$ schemes~\cite{Kuya_JCP_2018,Tamaki_JCP_2022}, by giving a more accurate reproduction of the induced entropy balance and by additionally enforcing the PEP property.

The enforcement of the PEP property has been seen as a crucial requirement in many applications~(\cite{Abgrall_JCP_2001,Shima_JCP_2021,Bernades_JCP_2023}) and recently it has been the subject of many studies, for ideal and real gases or multi-component flows (\cite{Shima_JCP_2021,Ranocha_CAMC_2021,Jain_JCP_2022,Fujiwara_JCP_2023,Bernades_JCP_2023}).
In this paper, we propose an analysis of existing PEP schemes for ideal gases which allows one to obtain simple criteria to enforce the PEP property in existing and widely used numerical fluxes. It is shown that the needed modifications typically amount to the specification of a suitable `dual' interpolation for the density in the internal-energy flux which, in the cases here analyzed, reduces to the use of the harmonic mean in place of the arithmetic mean.
The novel formulations are found to improve the global performances of the original schemes without increasing the computational cost.

\section{Problem formulation and discrete approximation} \label{sec:ProbForm}

In this paper, we focus on the spatial discretization of the compressible Euler equations:
\begin{align}
\dfrac{\partial \rho}{\partial t} &= -\dfrac{\partial \rho u_{\alpha}}{\partial x_{\alpha}} \;, \label{eq:Mass} \\[3pt]
\dfrac{\partial \rho u_{\beta}}{\partial t} &= -\dfrac{\partial \rho u_{\alpha}u_{\beta}}{\partial x_{\alpha}} -\dfrac{\partial p}{\partial x_{\beta}}  \;, \label{eq:Momentum} \\[3pt]
\dfrac{\partial \rho E}{\partial t} &=  -\dfrac{\partial \rho u_{\alpha}E}{\partial x_{\alpha}} -\dfrac{\partial p u_{\alpha}}{\partial x_{\alpha}},  \; \label{eq:TotEnergy}
\end{align}
where $\rho$, $u_{\alpha}$, $p$ and $E$ are respectively the density, the Cartesian velocity component along the direction $x_{\alpha}$, the pressure and the
total energy per unit mass, sum of internal and kinetic
energies: $E = e + u_{\alpha}u_{\alpha}/2$. 
We use Greek subscripts to refer to the Cartesian components of vectors, for which the summation convention over repeated indices holds.
When the Greek subscripts are omitted, it is assumed
that the relevant equations hold for a generic value of it (i.e.~$\alpha = 1,2$ or $3$).
In place of Eq.~\eqref{eq:TotEnergy}, the system of Euler equations could be equivalently written by using the evolution equation for another thermodynamic variable, such as pressure or internal energy per unit volume;
we discuss this possibility in the context of the discretization procedure in Sec.~\ref{sec:DiscrPressEq}.
The ideal gas law is assumed, which implies the simple proportionality $p = (\gamma-1)\rho e$,
the ratio of specific heats at constant pressure and volume $\gamma = c_p/c_v$
being constant.

We  work in a semi-discretized framework, in which 
spatial discretization is firstly performed, whereas the
resulting system of Ordinary Differential Equations (ODE) is integrated in time by
using a standard solver.
Since we focus on spatial discretization, we will assume that the effects of time integration errors are negligible at sufficiently small time steps.
This implies that in the theoretical analysis, all the
manipulations involving time derivatives can be carried out at the continuous level.
Spatial discretization is made by using a FD method over a uniform (colocated) Cartesian 
mesh of width $h$. 
Our notion of FD methods liberally includes also all the discretizations in which a numerical flux defined on the face located at the midpoint between nodal values is specified, and the convective terms are expressed as differences of numerical fluxes at adjacent faces. These formulations usually belong to FV methods, but in our use, the discrete variables are always interpreted as nodal values over a colocated mesh, and no reference to cell-vertex or cell-centered formulations is made.
Latin subscripts as $i,j$ or $k$ are used to denote the values of the discretized
variable on a nodal point $x_i$.

In what follows, we will discuss the PEP property of the spatial discretization;
for this, we need to analyze the 
induced evolution equations for $u_{\beta}$ and $p$, which can be easily obtained by manipulating Eq.~\eqref{eq:Mass}--\eqref{eq:TotEnergy} and read
\begin{align}
\dfrac{\partial u_{\beta}}{\partial t} &= -u_{\alpha}\dfrac{\partial u_{\beta}}{\partial x_{\alpha}} - \dfrac{1}{\rho}\dfrac{\partial p}{\partial x_{\beta}},\label{eq:Velocity} \\
\dfrac{\partial p}{\partial t} &=  -\dfrac{\partial p u_{\alpha}}{\partial x_{\alpha}} -\left(\gamma -1\right)p\dfrac{\partial u_{\alpha}}{\partial x_{\alpha}}.\label{eq:Pressure}
\end{align}
From these equations, one readily sees that for constant initial distributions of velocity and pressure, the right-hand sides of the evolution equations for velocity and pressure are zero, which in turn implies that time derivatives at the left-hand sides are also zero. This means that pressure and velocity remain constant (in time) when starting from constant (in space) distributions, i.e.~the equations preserve the equilibrium of the pressure and the solution evolves as a density wave, according to Eq.~\eqref{eq:Mass}.
An important point here is that Eq.~\eqref{eq:Velocity} and \eqref{eq:Pressure} are obtained by manipulation of Eqs.~\eqref{eq:Mass}--\eqref{eq:TotEnergy} in which the product or chain rules for derivatives are used, which are not valid, in general, at the discrete level.
This implies that when Eqs.~\eqref{eq:Mass}--\eqref{eq:TotEnergy} are directly discretized, the discrete analogues of Eq.~\eqref{eq:Velocity} and \eqref{eq:Pressure} have a structure which is, in general, different from that of the continuous equations, and the PEP property is typically lost at discrete level. 

Eqs.~\eqref{eq:Mass}--\eqref{eq:TotEnergy}, as most of the induced balance equations for secondary quantities, have a common structure that can be loosely expressed as
\begin{equation}\label{eq:GenStructBalEq}
    \dfrac{\partial \rho\phi}{\partial t}=-\mathcal{C}_{\rho\phi}-\mathcal{P}_{\rho\phi},
\end{equation}
where $\mathcal{C}_{\rho\phi}$ is the convective term
and $\mathcal{P}_{\rho\phi}$ is a pressure term.
The symbols $\mC$ and $\mP$ will be used here to denote both the individual spatial
terms at the right hand sides of Eqs.~\eqref{eq:Mass}--\eqref{eq:TotEnergy},
or their spatial discretizations,
the correct interpretation emerging from the context.
When needed, we will use the more explicit notation $\left.\mC_{\rho\phi}\right|_i$ to denote a particular discretization of $\mC_{\rho\phi}$ at node $x_i$.
In Eqs.~\eqref{eq:Mass}--\eqref{eq:TotEnergy} the convective terms have a divergence structure (i.e.~they are in the form
of the divergence of a convective flux) and we assume that the schemes used for the convective terms 
in the equations that are directly discretized are always in locally conservative form, i.e.~they can be expressed as a sum of differences of (numerical) fluxes along each Cartesian direction. We use the symbol $\mF^{\alpha}_{\rho\phi}$ to denote the numerical flux 
along the direction $x_{\alpha}$. With this notation, the discretization of the convective term $\mC_{\rho\phi}$ can be expressed as
\begin{equation}
\left.\mC_{\rho\phi}\right|_i = \dfrac{1}{h}\sum_{\alpha}\Delta\left.\mF^{\alpha}_{\rho\phi}\right|_{i}\label{eq:DiffOfFluxes}
\end{equation}
where $\Delta\mF_i = \mF_{i+1/2} - \mF_{i-1/2}$.
In most cases, the theory will be developed, without loss of generality, for the one-dimensional version of Eq.~\eqref{eq:Mass}--\eqref{eq:TotEnergy}, as the convective terms along the various directions are independent and can be treated separately. For this reason, the apex $\alpha$ will be most of the time removed from the notation for the numerical flux. Moreover, when the suffix $i+1/2$ or $i-1/2$ is not present, it is implicitly assumed that we are referring to the `right' flux: $\mF_{\rho\phi} = \left.\mF_{\rho\phi}\right|_{i+1/2}$.

As $\mC_{\rho\phi}$ is a divergence-type term, Eq.~\eqref{eq:DiffOfFluxes} suggests that $\mF_{\rho\phi}$ is an average- or interpolation-type term along $x$, consistent with 
the convective flux $\rho u\phi$ at $x_{i+1/2}$.
There are many ways in which this interpolation can be done, as one can choose to interpolate the product of the variables or to take the product of the interpolations. 
Moreover, the choice of the interpolation operator adds crucial degrees of freedom to the specification of the numerical flux.
Each choice corresponds to a different discrete formulation, with different properties. Here we are especially interested in how the choice of the numerical fluxes for the primary variables impacts the structural properties of the induced discrete evolution equations for pressure and velocity.

The formulations here analyzed will be exposed with respect to second-order (two-point), symmetric interpolations for the fluxes $\mF_{\rho\phi}$, which correspond to central-like and non-dissipative 
three-point discretizations of the convective terms $\mC_{\rho\phi}$.
However, provided that the interpolation operators are also smooth and consistent, 
the theory reported in \cite{DeMichele_JCP_2023} (and adapted from \cite{Ranocha_JSC_2018})
easily allows the extension to high-order formulations.

\section{Kinetic Energy preserving formulations}
\label{sec:KEP}
When using bilinear or trilinear interpolations, a complete correspondence can be obtained between the various choices of the numerical fluxes and suitable FD discretizations of the convective terms. For an extensive treatment of this and related topics, the reader is referred to \cite{Coppola_JCP_2019,Coppola_JCP_2023,Coppola_ECCOMAS_2022,DeMichele_C&F_2023,DeMichele_ECCOMAS_2022}. 
Here, we summarize only the most useful relations. 
We preliminarily define the arithmetic and product means as 
$\mean{\phi} = (\phi_{i}+\phi_{i+1})/2$ and 
$\pmean{\phi\psi} = (\phi_{i}\psi_{i+1}+\phi_{i+1}\psi_i)/2$ 
and the following FD discretizations for the convective terms for the mass equation and for a generic species $\phi$:
\begin{equation}\label{eq:Mass_Forms}
    \mathcal{M}^D = \dfrac{\delta \rho u }{\delta x },\qquad
    \mathcal{M}^A = \rho\dfrac{\delta u}{\delta x}+u\dfrac{\delta \rho}{\delta x}, \qquad
    \mM^{\textit{SKW}}  = (\mM^D + \mM^A)/2,
\end{equation}
\begin{equation}\label{eq:Mom_Forms}
    \mathcal{C}_{\rho\phi}^D = \dfrac{\delta \rho u \phi}{\delta x}, \quad
    \mathcal{C}_{\rho\phi}^{\phi} = \phi \dfrac{\delta \rho u}{\delta x} + \rho u\dfrac{\delta \phi}{\delta x},\quad
    \mathcal{C}_{\rho\phi}^{u} = u\dfrac{\delta \rho \phi}{\delta x}+\rho \phi\dfrac{\delta u}{\delta x},\quad
    \mathcal{C}_{\rho\phi}^{\rho}=\rho \dfrac{\delta u \phi}{\delta x} + \phi u\dfrac{\delta \rho}{\delta x}.
\end{equation}
where $\delta f = f_{i+1}-f_{i-1}$ is the central difference and $\mM$ is a special symbol we use to denote $\mC_{\rho}$.
With this notation, it is not difficult to recognize the following correspondences
between the convective terms and their associated fluxes in a 1D framework~\cite{Pirozzoli_JCP_2010,Coppola_JCP_2019,Coppola_JCP_2023,DeMichele_C&F_2023}
\begin{equation}\label{eq:InterpMass}
\mM^D  \longrightarrow  \mF_{\rho} = \mean{\rho u},\qquad
\mM^A  \longrightarrow  \mF_{\rho} = \pmean{\rho u},\qquad
\mM^{\textit{SKW}}  \longrightarrow  \mF_{\rho} = \mean{\rho}\mean{u}.
\end{equation}
\begin{equation}\label{eq:InterpPhi}
\mC_{\rho\phi}^{D}  \longrightarrow  \mF_{\rho\phi}= \mean{\rho u\phi},\quad
\mC_{\rho\phi}^F    \longrightarrow \mF_{\rho\phi}= \mean{\rho u}\mean{\phi},\quad
\mC_{\rho\phi}^C    \longrightarrow \mF_{\rho\phi}= \pmean{\rho u}\mean{\phi},\quad
\mC_{\rho\phi}^{\textit{KGP}}  \longrightarrow \mF_{\rho\phi}= \mean{\rho}\mean{u}\mean{\phi}.
\end{equation}
where $\mC^F = (\mC_{\rho\phi}^{D} + \mC_{\rho\phi}^{\phi})/2$, 
$\mC^C = (\mC_{\rho\phi}^{u} + \mC_{\rho\phi}^{\rho})/2$ and
 $\mC^{\textit{KGP}} = (\mC_{\rho\phi}^{F} + \mC_{\rho\phi}^{C})/2$
 are the discretizations of the convective terms used by \citet{Feiereisen_1981},
 \citet{Coppola_JCP_2019}, and \citet{Kennedy_JCP_2008} and 
 \citet{Pirozzoli_JCP_2010}, respectively.

 The coordinated use of suitable discretizations for the convective 
 terms in the equations for $\rho$ and $\rho\phi$ 
 can conduct to KEP
 methods,
 which are built in such a way that the convective term in the discrete evolution equation for the generalized kinetic energy $\rho\phi^2/2$ automatically has a globally (and locally) conservative structure~\cite{Coppola_AIMETA_2017,Coppola_AMR_2019}.
The combinations of the discretizations
$(\mM^D-\mC_{\rho\phi}^{F})$, $(\mM^A-\mC_{\rho\phi}^{C})$ and $(\mM^{\textit{SKW}}-\mC^{\textit{KGP}}_{\rho\phi})$ for the convective terms of the mass and $\phi$ equations are easily seen to be KEP~\cite{Coppola_JCP_2019,Coppola_JCP_2023,Coppola_AMR_2019}.
The same considerations hold for any linear 
combination of two of them, giving rise to a one-parameter family of locally conservative KEP schemes.
Its formulation in terms of numerical fluxes has the property that the convective fluxes for $\rho$ and $\rho\phi$ are linked by the relation $\mF_{\rho\phi} = \mF_{\rho}\mean{\phi}$, which assures kinetic-energy preservation in second-order, finite volume discretizations~\cite{Jameson_JSC_2008b,Veldman_JCP_2019}.
Note that this last condition is more general than the one developed for purely FD discretizations, as it holds for arbitrary specifications of the interpolations used for $\mF_{\rho}$. This observation is relevant for us, since in what follows we will use more general interpolations than the ones corresponding to classical FD formulations (e.g.~the logarithmic or harmonic means).
Moreover, the KEP condition expressed in terms of numerical fluxes precisely identifies the degree of freedom left in these formulations with the mass flux, which is specified by one parameter for two-point bilinear interpolations~\cite{Coppola_JCP_2023}, but can be arbitrarily specified for more general averages.
In all the cases analyzed, it can be shown that KEP schemes are both globally and locally conservative for the generalized kinetic energy $\rho\phi^2/2$, with convective numerical flux $\mF_{\rho\phi^2/2}=\mF_{\rho}\phi_i\phi_{i+1}/2$ (\cite{Coppola_JCP_2023,DeMichele_JCP_2023}).

\section{Pressure Equilibrium Preserving formulations}\label{sec:PEP_Form}
\subsection{Discrete velocity equation}
Starting from the (semi-)discrete equations \eqref{eq:GenStructBalEq}
for mass ($\phi=1$) and momentum ($\phi=u$), one can easily obtain the induced discrete version of the velocity equation~\eqref{eq:Velocity} 
\begin{equation}\label{eq:DiscrVelEvol}
\dfrac{\partial u}{\partial t} = \dfrac{u}{\rho} \mM-\dfrac{\mC_{\rho u}}{\rho}
-\dfrac{\mP_{\rho u}}{\rho}.
\end{equation}
To discretely preserve the pressure equilibrium, the right-hand side of this equation must be equal to zero when pressure and velocity are spatially constant.
Assuming that both $\mM$ and $\mC_{\rho u}$ do not explicitly depend on the pressure, one obtains the conditions
\begin{align}
\label{eq:PEP_condition_rhou_pressure}
    \hat{\mP}_{\rho u} &= 0\\
\label{eq:PEP_condition_rhou_convective}
    \hat{\mC}_{\rho u} &= U\hat{\mM}
\end{align}
where we use the notation that the `hat' symbol $\,\hat\,$ denotes the discrete term when evaluated at constant values of velocity $u = U$ and pressure $p = P$.

The condition in Eq.~\eqref{eq:PEP_condition_rhou_pressure} is always verified when a straightforward discretization of the pressure term is performed, such as $\mP_{\rho u} = \delta p/\delta x$. More elaborate discretizations for $\mP_{\rho u}$, however, are still allowed to obtain a PEP scheme, as long as Eq.~\eqref{eq:PEP_condition_rhou_pressure} is satisfied. 
The condition in Eq.~\eqref{eq:PEP_condition_rhou_convective} needs more attention. 
Expressed in flux form it requires \cite{Ranocha_CAMC_2021}
\begin{equation}\label{eq:PEP_condition_rhou_convective_flux}
    \hat{\mF}_{\rho u} = \hat{\mF}_{\rho} U + \mathop{\text{const}}.
\end{equation}
As already mentioned in Sec.~\ref{sec:KEP}, a generic KEP scheme in flux form satisfies $\mF_{\rho u} = \mF_{\rho}\mean{u}$, so this kind of scheme fulfills the condition in Eq.~\eqref{eq:PEP_condition_rhou_convective_flux} (and Eq.~\eqref{eq:PEP_condition_rhou_convective}). 
This is true for all schemes with mass flux based on bilinear interpolations, but also for more general schemes in which arbitrary interpolations are used for $\mF_{\rho}$. Moreover, Eq.~\eqref{eq:PEP_condition_rhou_convective} can be satisfied also by schemes that are not KEP. 
As an example, the formulation employed by \citet{Blaisdell_ANM_1996}, 
corresponding to $\mM = \mM^{\textit{SKW}}$ and $\mC_{\rho u} = \left(\mC_{\rho u}^{D}+\mC_{\rho u}^{u}\right)/2$
has numerical fluxes $\mF^{\alpha}_{\rho} = \mean{\rho}\mean{u}_{\alpha}$ and $\mF_{\rho u_{\beta}}^{\alpha} = \mean{\rho u}_{\beta}\,\mean{u}_{\alpha}$, which are easily seen to be not KEP.
However, for constant velocity and pressure, they become 
$\hat{\mF}_{\rho}^{\alpha}=\mean{\rho}U_{\alpha}$ and $\hat{\mF}_{\rho u_{\beta}}^{\alpha}=\mean{\rho}U_{\beta}U_{\alpha}$, which
 satisfy Eq.~\eqref{eq:PEP_condition_rhou_convective_flux}. 
It turns out that the formulation employed in~\citet{Blaisdell_ANM_1996} for the energy equation also satisfies the analogous condition for pressure, detailed in the next Sec.~\ref{sec:DiscrPressEq}, which shows that the Blaisdell scheme, even without being KEP, is PEP. 
 \subsection{Discrete pressure equation}\label{sec:DiscrPressEq}
As for the case of the discrete equation for velocity, algebraic manipulation of equations \eqref{eq:GenStructBalEq} for mass, momentum, and an 
`energy' variable among $\rho E$, $\rho e$ or $p$, allows one to write the discrete evolution equation for $p$. 
The form of this equation depends on the choice of the primary energy variable whose equation is directly discretized. 
In the following sections, we separately analyze some of the possible options.

\subsubsection{Formulations based on the pressure equation}
The simplest case is when the pressure equation \eqref{eq:Pressure} is discretized in place of Eq.~\eqref{eq:TotEnergy}.
In this case, one has that the discrete pressure equation trivially is
\begin{equation}\label{eq:Discr_p_Eq}
    \dfrac{\partial p}{\partial t}=-\mathcal{C}_{p}-\mathcal{P}_{p},
\end{equation}
where $\mC_{p}$ and $\mP_{p}$ are specified through direct discretization of the terms $\partial_x pu$ and $(\gamma-1)p\partial_x u$, respectively.
Assuming a straightforward discretization of the non-conservative pressure term, 
such as 
\begin{equation}\label{eq:Press_p_discr}
    \mP_{p}=\left(\gamma -1\right)p\dfrac{\delta u}{\delta x},
\end{equation} 
and of $\mC_p$, by using a general average of divergence and advective forms 
\begin{equation}
\mC_p = \chi\dfrac{\delta pu}{\delta x} + \left(1-\chi\right)\left(p\dfrac{\delta u}{\delta x} +u\dfrac{\delta p}{\delta x} \right), \label{eq:Conv_p_discr}
\end{equation}
one naturally has $\hat{\mC}_p = \hat{\mP}_p = 0$.
The simple proportionality between pressure and internal energy implies that if pressure is correctly preserved by convection (i.e.~$\mC_p$ is in locally conservative form), then so is internal energy. Moreover, if the discretization of density and momentum equations is such that the scheme is KEP, then total energy is also conserved as a result.

This approach has been followed for the first time by \citet{Feiereisen_1981}. 
In that paper, in addition to the well known KEP discretization of mass and momentum equations, they use the pressure equation as `energy' equation,
with convective term discretized using the advective form
\begin{equation}\label{eq:Feiereisen_p_FD}
    \mC_{p} = p\dfrac{\delta u}{\delta x} + u\dfrac{\delta p}{\delta x},
\end{equation}
corresponding to $\chi=0$ in Eq.~\eqref{eq:Conv_p_discr} and to the convective numerical flux $\mF_p=\pmean{pu}$.
The non-conservative pressure term is discretized as usual as $\mP_{p}=(\gamma-1)p\delta u/\delta x$.
Under these assumptions, 
the condition  $\hat{\mC}_p = \hat{\mP}_p = 0$ is easily verified, which assures that the Feiereisen scheme, which is a prototypical KEP scheme, is also PEP.
More recently, \citet{Shima_JCP_2021} proposed the KEEP-PE discretization of the total-energy equation that, for exact time integration, is equivalent to a direct discretization of the pressure equation where the skew-symmetric form is employed for the convective term 
\begin{equation}\label{eq:KEEP-PE_p_FD}
\mC_p = \frac{1}{2}\dfrac{\delta pu}{\delta x} + \frac{1}{2}\left(p\dfrac{\delta u}{\delta x} +u\dfrac{\delta p}{\delta x} \right) .
\end{equation}
corresponding to $\chi=1/2$ in Eq.~\eqref{eq:Conv_p_discr} with associated numerical flux $\mF_{p} = \mean{p}\mean{u}$.
The general formulation in Eq.~\eqref{eq:Conv_p_discr}  produces valid PEP schemes for all values of $\chi$
with numerical flux $\mF_{p} = \chi \mean{p u} + \left(1-\chi\right) \pmean{pu}$.
The case of $\chi=1$ has been also investigated in \cite{DeMichele_C&F_2023}.

Since the family defined by Eq.~\eqref{eq:Conv_p_discr} only includes schemes that are combination of advective and divergence forms, and as a consequence are based on bilinear fluxes, it excludes all the possible discretizations associated with nonlinear fluxes. 
In fact, in the more general framework of finite-volume methods, the convective term 
$\mC_p$ is built as a difference of arbitrary fluxes, whose generic form can be written as
\begin{equation}\label{eq:Conv_p_discr_F}
    \mF_p = \widetilde{pu}
\end{equation}
where $\widetilde{\psi\phi}$ represents a generic two-point average of the product $\psi \phi$.
Provided that the obvious consistency property that the flux is invariant under translation for constant $u$ and $p$ holds, one has $\left.\hat{\mF}_p\right|_{i-1/2} = \left.\hat{\mF}_p\right|_{i+1/2}$ and pressure equilibrium is preserved by convection.
An example of this more general case is given by
\begin{equation}\label{eq:Flux_pressure_geometric}
    \mF_p = \gmean{p}\mean{u},
\end{equation}
where $\gmean{\phi}$  is the geometric mean: $\gmean{\phi} = \sqrt{\phi_i\phi_{i+1}}$.
The convective term built on this flux can be shown to be equivalent to the discretization proposed by \citet{Rozema_JT_2014}. 
In their work, it is considered the evolution equation of the square-root variable $\sqrt{\rho e }$
which, due to the proportionality between $\rho e$ and $p$ for ideal gases, is equivalent to a direct discretization of the equation
\begin{equation}
    \frac{\partial \sqrt{p}}{\partial t} =
    -\dfrac{\partial \sqrt{p} u_{\alpha}}{\partial x_{\alpha}}    
    -\left(\frac{\gamma}{2} -1\right)\sqrt{p}\dfrac{\partial u_{\alpha}}{\partial x_{\alpha}}.
\end{equation}
The use of the skew-symmetric form on the convective term can be shown to lead back to the flux in Eq.~\eqref{eq:Flux_pressure_geometric}.

The findings of this section can be summarized by the following
\begin{remark}\label{remark1}
{\it (Discrete pressure equation)} \\
A finite-difference discretization of the pressure equation with formulations \eqref{eq:Conv_p_discr} (or \eqref{eq:Conv_p_discr_F}) for the convective term and \eqref{eq:Press_p_discr} for the pressure term, is always PEP, provided the conditions for velocity equilibrium Eq.~\eqref{eq:PEP_condition_rhou_convective} and \eqref{eq:PEP_condition_rhou_pressure} are satisfied. For ideal gases, internal energy is additionally discretely preserved by convection. If the discretizations of mass and momentum equations are KEP,  total energy is also discretely conserved.
\end{remark}

\subsubsection{Formulations based on the internal-energy equation}
 All the mentioned formulations based on a direct discretization of the pressure equation can also be interpreted as discretizations of the internal-energy equation
 \begin{equation} \label{eq:IntEnergy}
\dfrac{\partial \rho e}{\partial t} = -\dfrac{\partial \rho u_{\alpha} e}{\partial x_{\alpha}} -p\dfrac{\partial u_{\alpha}}{\partial x_{\alpha}},
\end{equation}
as, due to the proportionality relation in the ideal gas law, one has
 \begin{equation}\label{eq:ConvIntEn}
   \mC_{\rho e} = \frac{1}{\gamma-1}\mC_p,\qquad   \mP_{\rho e} = \frac{1}{\gamma-1}\mP_p,\qquad
     \mF_{\rho e} =\frac{1}{\gamma-1} \mF_p.
 \end{equation}
However, it is also possible to have schemes defined from a direct  discretization of the internal-energy equation \eqref{eq:IntEnergy} that cannot be interpreted as coming from a straightforward discretization of the pressure equation.
This happens when, in the specification of the numerical flux $\mF_{\rho e}$, the product $\rho e$ is not treated as one single variable, but $\rho$ and $e$ are averaged separately. 
In this case the discretization of Eq.~\eqref{eq:IntEnergy} is not equivalent to a direct discretization of the pressure equation,
as the product of interpolations of $\rho$ and $e$ does not correspond, in general, to a single interpolation involving only pressure.

An example of this type of schemes is the KGP$(\rho e)$ formulation analyzed in \citet{Coppola_JCP_2019}, which uses the internal energy as thermodynamic variable and discretizes the convective terms with a full splitting, using the forms 
$\mM^{\textit{SKW}}$ for the mass equation and $\mC^{\textit{KGP}}_{\rho\phi}$ for momentum and internal-energy equations, resulting in the set of fluxes
\begin{equation}\label{eq:KGPrhoe_fluxes}
\mathcal{F}_{\rho}= \mean{\rho}\,\mean{u},\qquad\qquad
\mathcal{F}_{\rho u} =\mathcal{F}_{\rho}\,\mean{u},\qquad\qquad
\mathcal{F}_{\rho e} = \mathcal{F}_{\rho}\,\mean{e}.
\end{equation}
For exact time integration, the scheme in Eq.~\eqref{eq:KGPrhoe_fluxes} is equivalent to the KEEP scheme proposed in~\citet{Kuya_JCP_2018}, which discretizes the total-energy equation with kinetic-energy and pressure fluxes consistent with that induced by the formulation in Eq.~\eqref{eq:KGPrhoe_fluxes}.
The  KEEP scheme is analyzed in the next Sec.~\ref{sec:total_energy_discretization}.
The scheme defined by Eq.~\eqref{eq:KGPrhoe_fluxes} cannot be cast as a conservative discretization of the pressure equation and, in fact, it is not PEP, as shown below.

Due to the proportionality relation in Eq.~\eqref{eq:ConvIntEn}, schemes based on the discretization of the internal-energy equation can be PEP if, for constant $p$ and $u$, the convective and pressure term for $\rho e$  fulfill the condition
\begin{equation}\label{eq:PEP_Cond_2}
\hat{\mC}_{\rho e} + \hat{\mP}_{\rho e}= 0
\end{equation}
as the proportionality condition $\partial_t\rho e=(\gamma-1)\partial_t p$ (inducing Eq.~\eqref{eq:ConvIntEn}) holds in all cases for ideal gases.
If the pressure term in Eq.~\eqref{eq:IntEnergy} is directly discretized as $p\delta u/\delta x$, one has $\hat{\mP}_{\rho e}=0$ and Eq.~\eqref{eq:PEP_Cond_2} is equivalent to requiring that $\hat{\mC}_{\rho e}=0$, i.e.~the flux for the convective part of the internal-energy equation is only a function of pressure and velocity, which implies $\hat{\mF}_{\rho e} = \mathop{\text{const}}$.
Since we have already considered the case in which the product $\rho e$ is treated as one, let us consider the generic internal-energy flux in which $\rho$ and $e$ are averaged separately. In this case, when velocity and pressure are constant and equal to $U$ and $P$ respectively, one can write:
\begin{equation}\label{eq:Flux_eint_upcost}
    \hat{\mF}_{\rho e} = \meanone{\rho}\,\meantwo{e} U =\meanone{\rho}\,\meantwo{\rho^{-1}} \frac{P U}{\gamma -1}
\end{equation}
in which two different kinds of mean have been used for density and internal energy. To obtain Eq.~\eqref{eq:Flux_eint_upcost} we have assumed that, given a constant $a$ and a variable $\phi$, the relation $\tildemean{a\phi} = a \tildemean{\phi}$ holds for a generic mean. This is not the case for every possible mean, but it holds for most of them, such as all power means, including the harmonic mean, the logarithmic mean and the Stolarsky means used in~\citet{Winters_BIT_2020}.

In order for the flux in Eq.~\eqref{eq:Flux_eint_upcost} to be constant, the means chosen for density and internal energy must satisfy
\begin{equation}\label{eq:condition_mean_PEP}
    \meanone{\phi}\,\meantwo{\phi^{-1}} = 1
\end{equation}
where the value of the arbitrary constant at the right-hand side has to be one for consistency.
Eq.~\eqref{eq:condition_mean_PEP}  is the general condition for PEP schemes which are based on a discretization of the internal-energy equation in which $\rho$ and $e$ are averaged separately.
The previous discussion can be summarized in the following
\begin{remark}\label{remark2}
{\it (Discrete internal-energy equation)} \\
For ideal gases, a locally-conservative discretization of the internal-energy equation in which the convective flux is specified treating the product of  density and internal energy as a single variable is equivalent to a formulation based on the pressure equation. As such, it is always PEP under the hypotheses of Remark~\ref{remark1}.
Locally-conservative discretizations in which density and internal energy are interpolated separately are PEP if Eq.~\eqref{eq:condition_mean_PEP} is satisfied.
\end{remark}

Many popular schemes fall into this category, 
and it is a simple exercise to check if they are PEP through the use of Eq.~\eqref{eq:condition_mean_PEP}.
The simplest example is the KGP$(\rho e)$ scheme defined in Eq.~\eqref{eq:KGPrhoe_fluxes} for which one has
\begin{equation}\label{eq:Flux_eint_KGP}
\mF_{\rho e} = \mean{\rho}\mean{u}\mean{e} \quad\longrightarrow\quad\hat{\mathcal{F}}_{\rho e} = \mean{\rho}\mean{\rho^{-1}} \frac{P U}{\gamma -1}.
\end{equation}
It is easily seen that in this case the flux is not constant, since 
 $\mean{\rho}\mean{\rho^{-1}} = \left(\mean{\rho}/\gmean{\rho}\right)^2$, which do not satisfy Eq.~\eqref{eq:condition_mean_PEP}, in general.

The scheme with flux in Eq.~\eqref{eq:Flux_pressure_geometric},  can also be reinterpreted as a formulation based on internal-energy equation. 
In fact, due to the identity $\gmean{\phi\psi} = \gmean{\phi}\gmean{\psi}$, the internal-energy flux corresponding to the pressure flux
in Eq.~\eqref{eq:Flux_pressure_geometric} is $\mathcal{F}_{\rho e} = \gmean{\rho}\gmean{e}\mean{u}$,
for which one has
\begin{equation}\label{eq:Flux_eint_geometric}
\hat{\mathcal{F}}_{\rho e} = \gmean{\rho}\gmean{\rho^{-1}} \frac{P U}{\gamma -1}.
\end{equation}
By noting that the geometric mean naturally has the property $1/\gmean{\phi}=\gmean{\phi^{-1}}$, one sees that 
Eq.~\eqref{eq:condition_mean_PEP} is satisfied with $\meanone{\phi}=\meantwo{\phi} = \gmean{\phi}$, which implies that the scheme based on the geometric mean, and equivalent to that proposed by~\citet{Rozema_JT_2014}, is PEP.

As a further example, we consider the scheme studied by 
\citet{Ranocha_CAMC_2021}, which can be formulated through the 
specification of the mass, momentum and internal-energy fluxes as~\cite{DeMichele_JCP_2023}
\begin{equation}\label{eq:Ranocha_Flux_eint}
\mathcal{F}_{\rho}= \overline{\rho}^{\text{log}}\,\overline{u},\qquad\qquad
\mathcal{F}_{\rho u} =\mathcal{F}_{\rho}\,\overline{u},\qquad\qquad
\mathcal{F}_{\rho e} = \mathcal{F}_{\rho}\,\left[\overline{\left(1/e\right)}^{\text{log}}\right]^{-1}
\end{equation}
where $\logmean{\phi} = (\rho_{i+1}-\rho_i)/(\log\rho_{i+1}-\log\rho_i)$ is the logarithmic mean.
This scheme is easily seen to be KEP, because of the form of the momentum flux, and the wise use of the logarithmic 
mean renders it also EC 
for perfect gases~\cite{Ranocha_CAMC_2021,DeMichele_JCP_2023}.
Written explicitly, the internal-energy flux is
\begin{equation}\label{eq:Ranocha_Flux_eint_explicit}
\mathcal{F}_{\rho e} = \logmean{\rho} \mean{u}\,\left[\logmean{\left(1/e\right)}\right]^{-1}
\end{equation}
for which
\begin{equation}\label{eq:Flux_Ranocha_Pconst}
\hat{\mathcal{F}}_{\rho e} = \logmean{\rho}\left[\logmean{\rho}\right]^{-1} \frac{P U}{\gamma -1} = \text{const.}
\end{equation}
and in this case $\meanone{\phi} = \logmean{\phi}$ and
$\meantwo{\phi}$ is taken exactly as the dual mean required to satisfy Eq.~\eqref{eq:condition_mean_PEP}:
$\meantwo{\phi}=\left[\logmean{\left(1/\phi\right)}\right]^{-1}$.
\

Finally, we consider the formulation recently proposed by \citet{DeMichele_JCP_2023}, whose convective fluxes for mass, momentum and internal-energy are
\begin{equation}\label{eq:Flux_ArithHarmonic}
\mathcal{F}_{\rho}= \mean{\rho}\,\overline{u},\qquad\qquad
\mathcal{F}_{\rho u} =\mathcal{F}_{\rho}\,\mean{u},\qquad\qquad
\mathcal{F}_{\rho e} = \mathcal{F}_{\rho}\hmean{e},
\end{equation}
where $\hmean{\phi} = \phi_i\phi_{i+1}/(\phi_i+\phi_{i+1})$ is the harmonic mean.
For constant $u$ and $p$ the internal-energy flux becomes $\hat{\mathcal{F}}_{\rho e} = \mean{\rho}\hmean{\rho^{-1}}PU/(\gamma -1)$ and, 
based on the fact that the harmonic mean is the dual of the arithmetic mean with respect to Eq.~\eqref{eq:condition_mean_PEP}, i.e.~$\left(\mean{\phi}\right)^{-1} = \hmean{\phi^{-1}}$, 
it follows that also the scheme 
in Eq.~\eqref{eq:Flux_ArithHarmonic}
is PEP. 
It is interesting to note that this property extends also to the whole family of Asymptotically Entropy Conserving (AEC) schemes presented 
in~\cite{DeMichele_JCP_2023}, in which the means of density and internal energy include truncated asymptotic expansions.
For these schemes the internal-energy convective flux is written as
\begin{equation}\label{eq:Flux_eint_expansion}
\mathcal{F}_{\rho e} = \left(\dfrac{\mean{\rho}}{\sum_{n=0}^N \frac{\langle\rho\rangle^{2n}}{2n+1}}\right)\mean{u}\left(\hmean{e}\sum_{n=0}^N \frac{\langle e\rangle^{2n}}{2n+1}\right) = \mean{\rho}_{\text{\!\!\tiny{AEC}\,}} \mean{u} \hmean{e}_{\text{\!\tiny{AEC}}},
\end{equation}
where $\langle \phi\rangle = (\phi_{i+1}-\phi_i)/\mean{\phi}$.
It can be verified that condition \eqref{eq:condition_mean_PEP} holds for every value of $N$, making each scheme of the family PEP.

 \subsubsection{Formulations based on the total-energy equation}\label{sec:total_energy_discretization}
 If a KEP discretization is used for the density and momentum equations, every direct discretization of internal-energy or pressure equations with locally conservative convective terms can also be interpreted as a conservative discretization of the total-energy equation. 
 In this case, total energy evolves driven by numerical fluxes which are the sum of the primary internal-energy flux and 
 the kinetic-energy flux induced by the discretization of mass and momentum:
 \begin{equation}\label{eq:TotEnFlux}
     \mF_{\rho E} = \mF_{\rho e} + \mF_{\rho} \frac{u_i u_{i+1}}{2},
 \end{equation}
 where $\mF_{\rho e}=\mF_{p}/(\gamma -1)$ in case the pressure equation is directly discretized.
 The pressure term $\mP_{\rho E}$ results discretized with the advective form 
 \begin{equation}\label{eq:TotEnPress}
     \mP_{\rho E} = p\frac{\delta u}{\delta x}+ u\frac{\delta p}{\delta x}
 \end{equation} 
 coming from the separate discretizations of the pressure terms in the momentum equation ($\delta p/\delta x$) and in the internal-energy (or pressure) equation ($p\delta u/\delta x$).
 These formulations can be directly designed starting from the total-energy equation, with numerical fluxes specified according to Eq.~\eqref{eq:TotEnFlux} and \eqref{eq:TotEnPress},
 and are equivalent, for exact time integration, to the formulations based on the flux $\mF_{\rho e}$; as such, the total-energy flux inherits 
 the PEP properties of the flux  $\mF_{\rho e}$.

An example of this type of formulations is the KEEP scheme proposed by~\citet{Kuya_JCP_2018}, which is defined by the set of mass, momentum, and total-energy fluxes given by:
\begin{equation}\label{eq:KEEP_Flux_Etot}
\mathcal{F}_{\rho}= \mean{\rho}\,\mean{u},\qquad\qquad
\mathcal{F}^*_{\rho u} =\mathcal{F}_{\rho}\,\mean{u} + \mean{p},\qquad\qquad
\mathcal{F}^*_{\rho E} = \mathcal{F}_{\rho}\,\mean{e} + \mathcal{F}_{\rho}\,\frac{u_iu_{i+1}}{2} + \pmean{pu}
\end{equation}
where the apex $^*$ is used to indicate that the flux includes also the contribution coming from the (conservative) pressure term.
The form of the total-energy flux was determined in~\citet{Kuya_JCP_2018} by satisfying the so-called {\it Analytical Relations}, which were introduced to improve the entropy conservation of the overall formulation. The final result of their analysis is that the specification of the kinetic-energy flux and of the pressure term in the total-energy flux has to be the same as that induced by the discretization of mass and momentum alone, which renders the formulation equivalent, for exact time integration, to a formulation based on the internal-energy equation. 

 It is also possible to have formulations based on a direct discretization of the total-energy equation, which do not come from formulations in terms of internal energy or pressure.
 An example of this type of discretization is given by the formulation used by Jameson~\cite{Jameson_JSC_2008b} and Pirozzoli \cite{Pirozzoli_JCP_2010}.
 In this scheme, the total-energy equation is written as
 \begin{equation}\label{eq:TotEn_Enthalpy}
     \dfrac{\partial\rho E}{\partial t} = -\dfrac{\partial\rho u H}{\partial x}
 \end{equation}
where $H = E+p/\rho$ is the total enthalpy per unit mass. 
By following the same convention introduced in Eq.~\eqref{eq:KEEP_Flux_Etot}, the convective term in Eq.~\eqref{eq:TotEn_Enthalpy} is here denoted as $\mC_{\rho E}^* = \mC_{\rho E} + \mP_{\rho E}$, where $\mP_{\rho E}$ is the conservative pressure term $\partial pu/\partial x$.
In the Jameson-Pirozzoli (JP) scheme, $\mC_{\rho E}^*$
is split with a fully triple (KGP) splitting, corresponding to the numerical flux $\mF^*_{\rho E}=\mean{\rho}\mean{u}\mean{H}$. The contextual use of the KEP discretization 
 $(\mM^{\textit{SKW}}-\mC^{\textit{KGP}}_{\rho\phi})$ leads
 to a formulation that can be expressed in numerical-flux terms as
\begin{equation}\label{eq:Flux_JP}
\mathcal{F}_{\rho}= \mean{\rho}\,\mean{u},\qquad\qquad
\mathcal{F}^*_{\rho u} =\mathcal{F}_{\rho}\,\mean{u} + \mean{p},\qquad\qquad
\mathcal{F}^*_{\rho E} = \mathcal{F}_{\rho}\mean{H}.
\end{equation}
By expanding the total-energy flux into its components one has
\begin{equation}
    \mF_{\rho E}^* = \mF_{\rho}\mean{e}+\mF_{\rho}\mean{\frac{u^2}{2}} + \mF_{\rho}\mean{\left(\frac{p}{\rho}\right)}
\end{equation}
from which one sees that the scheme cannot be obtained from a straightforward discretization of the internal-energy equation,
as the convective flux of kinetic energy is different from that induced by the KEP formulation ($\mF_{\rho}\mean{u^2}/2$ versus $\mF_{\rho}u_iu_{i+1}/2$).
In this case, the internal energy discretely evolves with a convective flux $\mF_{\rho e} =\mF_{\rho E} - \mF_{\rho u^2/2}$ which reads
\begin{equation}
    \mF_{\rho e} =\mF_{\rho}\mean{e} + \mF_{\rho}\left(\frac{\mean{u^2}}{2}-\frac{u_iu_{i+1}}{2}\right).
\end{equation}
The (non conservative) pressure term is given by $\mP_{\rho e}= \mP_{\rho E} - u\,\delta p/\delta x$, i.e.
\begin{equation}
    \mP_{\rho e} =\Delta\left[\mF_{\rho}\mean{\left(\frac{p}{\rho}\right)}\right] -u \frac{\delta p}{\delta x}.
\end{equation}
For constant $u$ and $p$ one has
\begin{align}
    \hat{\mF}_{\rho e} &=\mean{\rho}\mean{\rho^{-1}}\frac{PU}{\gamma -1}\label{eq:PEPCond1_JP} \\
    \hat{\mP}_{\rho e} &=\Delta\left(\mean{\rho}\mean{\rho^{-1}}PU\right) \label{eq:PEPCond2_JP}.
\end{align}
The PEP condition now is that the right-hand side of Eq.~\eqref{eq:PEPCond1_JP} is constant and that 
of Eq.~\eqref{eq:PEPCond2_JP} is zero, neither of which is true, because of the term $\mean{\rho}\mean{\rho^{-1}}$, which shows that the JP scheme is not PEP.

A similar approach is adopted in the scheme proposed by~\citet{Singh_ArXiv_2021}, which in our notation can be written as
\begin{equation}\label{eq:Flux_Singh}
\mathcal{F}_{\rho}= \mean{\rho}\,\mean{u},\qquad\qquad
\mathcal{F}^*_{\rho u} =\mathcal{F}_{\rho}\,\mean{u} + \mean{p},\qquad\qquad
\mF_{\rho E}^* = \mean{\rho e}\mean{u} + \frac{1}{2}\mF_{\rho}\mean{u^2} +\mean{u}\mean{p}.
\end{equation}
In this case the internal-energy flux is split by grouping $\rho$ and $e$, which allows one to write the total-energy convective and pressure flux as
\begin{equation}\label{eq:Flux_SC}
\mF_{\rho E}^* = \frac{\gamma}{\gamma-1}\mean{p}\mean{u} + \mF_{\rho}\frac{\mean{u^2}}{2}.
\end{equation}
In this instance, however, the fulfillment of the PEP property is evident by observing that Eq.~\eqref{eq:Flux_Singh} induces
\begin{equation}\label{eq:Flux_Singh2}
\mF_{\rho e}= \mean{p}\mean{u} + \mF_{\rho}\left(\frac{\mean{u^2}}{2}-\frac{u_iu_{i+1}}{2}\right),\qquad\qquad
\mP_{\rho e} = \Delta \left(\mean{p}\mean{u}\right)-u\frac{\delta p}{\delta x}
\end{equation}
for which the conditions $\hat{\mF}_{\rho e} = \text{const.}$
and $\hat{\mP}_{\rho e} = 0$ are easily verified.

The discussion and examples above lead to the following
\begin{remark}
{\it (Discrete total-energy equation)} \\
A locally-conservative discretization of the total-energy equation,
that cannot be written as induced by an analogous formulation based on the internal-energy equation, is PEP provided that the kinetic-energy flux is constant  for constant velocity and pressure distributions and Eq.~\eqref{eq:condition_mean_PEP} is satisfied. For formulations induced by a discretization of the internal-energy equation the same conclusions of Remark~\ref{remark2} apply.
\end{remark}

\section{Novel PEP schemes}
The analysis exposed in the previous sections allows us to discuss the modifications needed to enforce the PEP property in existing schemes.
The general strategy is to express the given scheme in terms of fluxes for mass, momentum, and internal energy and to enforce the condition in Eq.~\eqref{eq:condition_mean_PEP} by modifying the interpolation for density in the internal-energy flux.
Here we consider two popular schemes that have been already analyzed in the previous sections, namely the KGP$(\rho e)$ (or KEEP) scheme in Eq.~\eqref{eq:KGPrhoe_fluxes} and \eqref{eq:KEEP_Flux_Etot} and the JP scheme in Eq.~\eqref{eq:Flux_JP}.

In the first case, inspection of Eq.~\eqref{eq:KGPrhoe_fluxes} shows that the internal-energy flux is $\mF_{\rho e} = \mean{\rho}\mean{u}\mean{e}$. 
For this scheme, one has $\meantwo{e} = \mean{e}$ and the modifications needed to render it PEP is the use of the dual of the arithmetic mean 
for density, which is the harmonic mean: $\meanone{\rho}=\hmean{\rho}$. In fact, the internal energy flux $\mF_{\rho e} = \hmean{\rho}\mean{u}\mean{e}$ reduces, for $u$ and $p$ constant to
 $\hat{\mF_{\rho e}} = \hmean{\rho}\mean{\rho^{-1}}PU/(\gamma -1)$, which is constant, since $\hmean{\rho}\mean{\rho^{-1}}=1$.
 The final expression for the modified KGP$(\rho e)$ scheme is
\begin{equation}\label{eq:KEEP-H1}
\mathcal{F}_{\rho}= \mean{\rho}\,\mean{u},\qquad\qquad
\mathcal{F}_{\rho u} =\mathcal{F}_{\rho}\,\mean{u},\qquad\qquad
\mF_{\rho e} = \hmean{\rho}\mean{u}\mean{e}.
\end{equation}
The discrete formulation based on this set of fluxes has the KEP and PEP properties and, as it will be shown in the next Sec.~\ref{sec:Results}, it basically retains the favorable properties on global entropy conservation of the original scheme.
Note that the scheme in Eq.~\eqref{eq:KEEP-H1} does not have the property that the internal-energy flux is the product between the mass flux and the arithmetic average for $e$ (as in the original version in Eq.~\eqref{eq:KEEP_Flux_Etot}), which is considered as a desirable characteristic since it adds the structural property that the quantity $\rho e^2/2$ is preserved globally (and locally) by convection.
To retain this property we propose to refine the formulation by adopting the harmonic mean also for the density in the mass flux, which leads to 
\begin{equation}\label{eq:KEEP-H}
\mathcal{F}_{\rho}= \hmean{\rho}\,\mean{u},\qquad\qquad
\mathcal{F}_{\rho u} =\mathcal{F}_{\rho}\,\mean{u},\qquad\qquad
\mF_{\rho e} = \mathcal{F}_{\rho}\mean{e}.
\end{equation}
This last scheme is the most straightforward modification of the KGP$(\rho e)$ or KEEP scheme which fulfills the PEP property, leaving as unmodified the general structure of the formulation. It will be referred to as the KGP$(\rho e)$-H (or KEEP-H in its formulation based on the total-energy flux). 
It can be also considered as a `dual' scheme with respect to the formulation defined in Eq.~\eqref{eq:Flux_ArithHarmonic} and presented in~\citet{DeMichele_JCP_2023}, since the means for density and internal energy in Eq.~\eqref{eq:Flux_ArithHarmonic} (arithmetic and harmonic, respectively) are exchanged in the scheme in Eq.~\eqref{eq:KEEP-H}.

It is interesting to note that, as it has been done for the AEC schemes in~\citet{DeMichele_JCP_2023}, the KGP$(\rho e)$-H scheme in Eq.~\eqref{eq:KEEP-H} can be 
made AEC through the use of truncated asymptotic expansions, still keeping the PEP property. 
In fact, the formulation given by 
\begin{equation}\label{eq:Flux_eint_keepn_hdensity}
\mathcal{F}_{\rho}= \left[\hmean{\rho}\dfrac{\sum_{n=0}^N \langle{\rho}\rangle^{2n}}{\sum_{n=0}^N \frac{\langle{\rho}\rangle^{2n}}{2n+1}}\right]\mean{u}=\hmean{\rho}_{\text{\!\tiny{AEC}}\,}\mean{u},\qquad\quad
\mathcal{F}_{\rho u} =\mathcal{F}_{\rho}\,\mean{u},\qquad\quad
\mathcal{F}_{\rho e} = \mathcal{F}_{\rho}\left[\mean{e}\dfrac{\sum_{n=0}^N \frac{\langle{e}\rangle^{2n}}{2n+1}}{\sum_{n=0}^N \langle{e}\rangle^{2n}}\right]=\massflux\,\mean{e}_{\!\!\text{\tiny{AEC}}}
\end{equation}
is easily shown to be Asymptotically Entropy Conservative, since the mass and internal-energy fluxes converge to the fluxes of Ranocha in Eq.~\eqref{eq:Ranocha_Flux_eint} as $N\rightarrow\infty$. The form of the truncated expansions in the mass and internal-energy fluxes is also such that the PEP property is retained at each order $N$. We will refer to this class of schemes as KGP$(\rho e)$-H$^{(N)}$ or KEEP-H$^{(N)}$.

A similar modification can be used to convert the JP scheme into a PEP scheme without altering the general structure of the fluxes. 
Once again, the failure to fulfill the PEP property in the JP scheme can be traced back to the term $\mean{\rho}\mean{\rho^{-1}}$ in the fluxes in Eqs.~\eqref{eq:PEPCond1_JP} and \eqref{eq:PEPCond2_JP}. The modification of the mass flux from $\mean{\rho}\,\mean{u}$ to $\hmean{\rho}\,\mean{u}$ converts the product $\mean{\rho}\mean{\rho^{-1}}$ in Eqs.~\eqref{eq:PEPCond1_JP} and \eqref{eq:PEPCond2_JP} into the (constant) product $\hmean{\rho}\mean{\rho^{-1}}$, which assures that the scheme is PEP.
Following these considerations, the proposed PEP modification of the JP scheme is defined by the fluxes
\begin{equation}\label{eq:JP-H}
    \mF_{\rho} = \hmean{\rho} \mean{u}\qquad\qquad
    \mF_{\rho}^* = \mF_{\rho} \mean{u} + \mean{p}\qquad\qquad
    \mF_{\rho E}^* = \mF_{\rho} \mean{H}
\end{equation}
We will refer to the formulation based on these fluxes as the JP-H scheme.

\section{Numerical results}\label{sec:Results}
In this section, two numerical tests are performed to assess the theoretical predictions and to test that the main properties of the original schemes are retained by the modified versions, which have the additional structural benefit of satisfying the PEP property.
In Tab.~\ref{tab:ConvProp} a list of the schemes used for the tests is reported, together with their conservation properties.
Independently of the energy variable they are based on, all schemes have been implemented through a discretization of total energy, so the total-energy flux is reported in Tab.~\ref{tab:ConvProp}. The high-order flux is obtained from the second-order, two-point one as explained in the appendix of \cite{DeMichele_JCP_2023}.
 \begin{table}
\renewcommand\arraystretch{1.8}
\centering
\begin{tabular}{ccccccccc}
&Ref.&&$\mF_{\rho}$ &   $\mF_{\rho E}^*$ & & KEP  & PEP  & AEC \\
\hline
KEEP & \cite{Kuya_JCP_2018} &&$ \mean{\rho}\mean{u}$ & $\massflux\,\mean{e} + \massflux\,\frac{u_iu_{i+1}}{2} + \pmean{pu}$& &$\checkmark$& $\times$ & $\times$ \\
KEEP-H & new &&$ \hmean{\rho}\mean{u}$ & $\massflux\,\mean{e} + \massflux\,\frac{u_iu_{i+1}}{2} + \pmean{pu}$ & &$\checkmark$& $\checkmark$ & $\times$ \\
KEEP-H${}^{(N)}$ & new &&$ \hmean{\rho}_{\text{\!\tiny{AEC}}\,}\mean{u}$& $\massflux\,\mean{e}_{\!\!\text{\tiny{AEC}}}+ \massflux\,\frac{u_iu_{i+1}}{2} + \pmean{pu}$ & &$\checkmark$& $\checkmark$ & $\checkmark$ \\
JP & \cite{Jameson_JSC_2008b},\cite{Pirozzoli_JCP_2010} &&$ \mean{\rho}\mean{u}$ & $\massflux\,\mean{e} + \massflux\,\frac{\mean{u^2}}{2} + \massflux\mean{\left(\frac{p}{\rho}\right)}$ & &$\checkmark$& $\times$ & $\times$ \\
JP-H & new &&$ \hmean{\rho}\mean{u}$ & $\massflux\,\mean{e} + \massflux\,\frac{\mean{u^2}}{2} + \massflux\mean{\left(\frac{p}{\rho}\right)}$ & &$\checkmark$& $\checkmark$ & $\times$ \\
$A\rho$-$He$& \cite{DeMichele_JCP_2023} &&$ \mean{\rho}\mean{u}$ & $\massflux\,\hmean{e} + \massflux\,\frac{u_iu_{i+1}}{2} + \pmean{pu}$& &$\checkmark$& $\checkmark$ & $\times$ \\
AEC${}^{(N)}$ & \cite{DeMichele_JCP_2023} &&$ \mean{\rho}_{\text{\!\!\tiny{AEC}\,}}\mean{u} $ & $\massflux\,\hmean{e}_{\text{\!\tiny{AEC}}} + \massflux\,\frac{u_iu_{i+1}}{2} + \pmean{pu}$& &$\checkmark$& $\checkmark$ & $\checkmark$ \\
\hline
\end{tabular}
\caption{
Fluxes and conservation properties of the various formulations considered. The definition of the means $\hmean{\rho}_{\text{\!\tiny{AEC}}\,}$ and $\mean{e}_{\!\!\text{\tiny{AEC}}}$ can be obtained from Eq.~\eqref{eq:Flux_eint_keepn_hdensity}; the meaning of $\mean{\rho}_{\text{\!\!\tiny{AEC}\,}}$, $\hmean{e}_{\text{\!\tiny{AEC}}}$ can be deduced from Eq.~\eqref{eq:Flux_eint_expansion}.
$\checkmark$: variable preserved locally and globally,
$\times$: variable not preserved.
 \label{tab:ConvProp}}
\end{table}

The first test is a simple two-dimensional density-wave problem, in which an initially constant (two--dimensional) distribution of pressure and velocity is assumed with a space-variable density field. The analytical solution evolves as a convection of the density wave, meanwhile both pressure and velocity are supposed to remain constant in time.
The initial conditions for the density wave test are
\begin{equation*}
    \rho_0 = 1 + \exp\left(\sin\left(2\pi (x+y)\right)\right), \qquad u_0 = 1, \qquad v_0 = 2, \qquad p_0 = 1, \qquad (x,y)\in[-1,1]^2.
\end{equation*}
The fluxes employed are fourth-order accurate  while the classical fourth-order Runge-Kutta scheme (RK4) is used for time integration.
The domain is discretized using $30$ nodes in each spatial direction; the time step is chosen so that $\text{CFL}=0.01$ and the error due to the temporal integration can be considered negligible.

\begin{figure}[tb]
    \centering
     \begin{subfigure}[b]{0.47\textwidth}
         \centering
         \includegraphics[width=\textwidth]{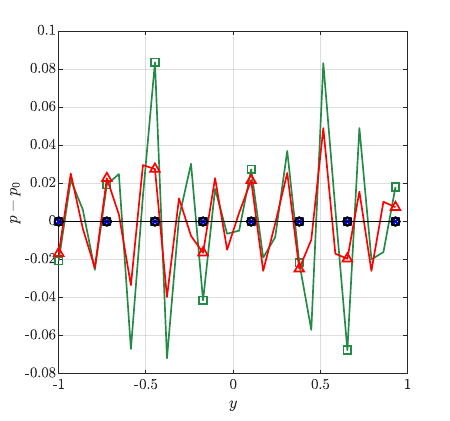}
         \caption{Pressure oscillation}
         \label{fig:pressure_DW}
     \end{subfigure}
     \begin{subfigure}[b]{0.47\textwidth}
         \centering
         \includegraphics[width=\textwidth]{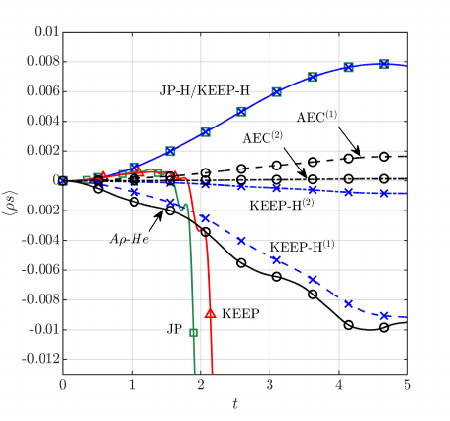}
         \caption{Entropy production}
         \label{fig:entropy_integral_DW}
     \end{subfigure}
    \caption{Density wave simulation using different numerical fluxes.
    On the left, comparison of the pressure solution at $x=0$ and at time $t=1.5$;
    on the right, time evolution of entropy integral.
    Black continuous lines with circles represent the $A\rho$-$He$ scheme, while dashed and dash-dotted lines are used for AEC${}^{(1)}$ and AEC${}^{(2)}$; red continuous lines with triangles represent the KEEP scheme; solid green with squares identifies the JP scheme and dashed is for JP-H; solid blue with cross signs is used for the KEEP-H flux, while dashed and dash-dotted lines are used for KEEP-H${}^{(1)}$ and KEEP-H${}^{(2)}$.
    The mesh is discretized in $30\times30$ nodes and $\textrm{CFL} = 0.01$.}
    \label{fig:DW}
\end{figure}

As expected from the theoretical analysis and from the results of previous studies \cite{Shima_JCP_2021,DeMichele_JCP_2023}, the two schemes lacking the PEP property, the KEEP and the JP schemes, induced spurious pressure oscillations, whose amplification lead to the blow-up of the simulations at times $2.5$ and $2.3$, respectively.
The error on pressure 
is showcased in Fig.~\ref{fig:pressure_DW}: at a time $t=1.5$, before the blow-up of any of the simulations, the PEP schemes have a pressure value that is within $2\times10^{-13}$ of the initial value, while the JP and KEEP schemes present nonphysical oscillations.
The error on entropy conservation is also investigated, to ensure that the modification that has been used to make the JP and KEEP schemes PEP did not drastically worsen their performance.
Fig.~\ref{fig:entropy_integral_DW} shows the time evolution of the entropy integral $\langle \rho s \rangle$ nondimensionalized as 
$\langle \phi \rangle =
\left.
 \left(\int_{\Omega}  \phi d \Omega - \int_{\Omega} \phi_0 d \Omega\right)\right/ \left(\left\lvert\int_{\Omega} \phi_0 d \Omega\right\rvert\right)$.
 In this test, the new schemes JP-H and KEEP-H have identical entropy production; this can be explained since, for constant $u$ and $p$, they both reduce to the scheme with density and internal energy fluxes as
\begin{equation*}
    \hat{\mF}_{\rho} =\hmean{\rho}U,\qquad
    \hat{\mF}_{\rho e} =\frac{PU}{\gamma -1}.
\end{equation*}
The comparison with the original JP and KEEP schemes is hard to draw, since they both quickly diverge, but they also seem to have a similar behavior, at least at the beginning.
The entropy error of the simulation using the scheme A$\rho$-H$e$ has the same order of magnitude, since it also has an identical internal energy flux to that of JP-H and KEEP-H for constant $u$ and $p$, and differs only in the specification of the mass flux ($\mean{\rho}U$ versus $\hmean{\rho}U$).
Larger differences can be observed by comparing the family of asymptotically entropy conserving KEEP-H${}^{(N)}$ and AEC${}^{(N)}$:
by increasing the number of additional terms $N$, the AEC${}^{(N)}$ family seems to be converging faster to an exactly entropy conserving flux, with $\langle\rho s\rangle\approx 1\times10^{-4} $ for AEC${}^{(2)}$ at $t=5$.
This is in accordance with the theory presented in \cite{DeMichele_JCP_2023}, by which the arithmetic mean on density and the harmonic mean on internal energy are usually closer to the means used by the entropy conserving scheme by Ranocha~\cite{Ranocha_JSC_2018}.

For the second test, all the schemes here considered have been implemented in the open-source code STREAmS-2~\cite{Bernardini_CPC_2023},  
which is a parallel, high-order compressible flow solver with GPU support.
The solver has been used to simulate the three-dimensional case of the Taylor-Green vortex, with initial conditions 
\begin{align*}
    \rho(x,y,z) &= \rho_0\\
    u(x,y,z) &= u_0\sin(x)\cos(y)\cos(z)\\
    v(x,y,z) &= -u_0\cos(x)\sin(y)\cos(z)\\
    w(x,y,z) &= 0\\
    p(x,y,z) &= p_0 + u_0^2\frac{(\cos(2x) + \cos(2y))(2+\cos(2z))}{16}
\end{align*}
in which $\rho_0=1$, $p_0=100$, $u_0=M \sqrt{\gamma p_0/\rho_0}$, with the Mach number $M = 0.2$.
The triperiodic domain has side length $2\pi$ in all directions and is discretized using $32\times32\times32$ nodes. 
For the spatial discretization, the 6th-order version of the fluxes is used, whereas for time integration a standard RK4 procedure has been implemented and used at CFL = 0.1, which is sufficiently small that linear invariants are exactly conserved to machine precision for all schemes. 
\begin{figure}[tb]
    \centering
     \begin{subfigure}[b]{0.47\textwidth}
         \centering
         \includegraphics[width=\textwidth]{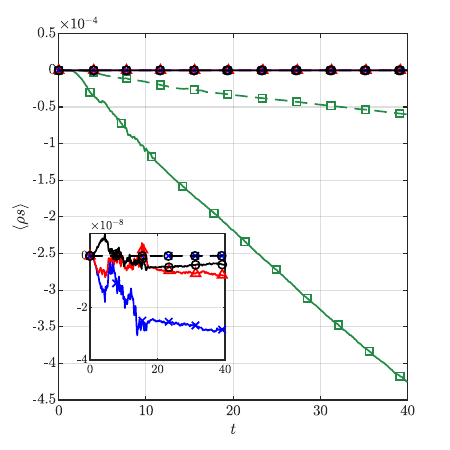}
         \caption{Entropy production}
         \label{fig:entropy_integral_TGV}
     \end{subfigure}
     \begin{subfigure}[b]{0.47\textwidth}
         \centering
         \includegraphics[width=\textwidth]{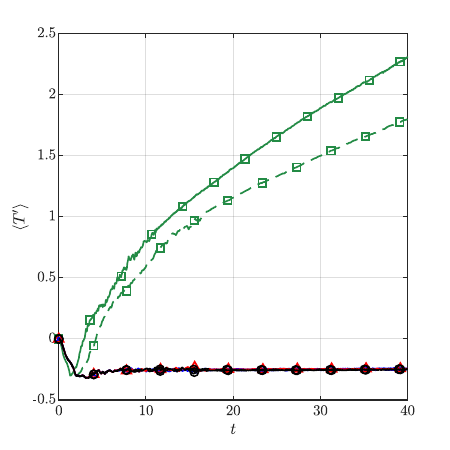}
         \caption{Temperature fluctuations}
         \label{fig:fluc_T_TGV}
     \end{subfigure}
    \caption{Taylor-Green vortex simulation at $M=0.2$ using different numerical fluxes. On the left, time evolution of entropy integral:
    black continuous lines with circles represent the $A\rho$-$He$ scheme, while dashed lines are used for AEC${}^{(1)}$; red continuous lines with triangles represent the KEEP scheme; solid green with squares identifies the JP scheme and dashed is for JP-H; solid blue with cross signs is used for the KEEP-H flux, while dashed lines are used for KEEP-H${}^{(1)}$. On the right, the evolution of the temperature fluctuations.
    The mesh is discretized in $32\times32\times32$ nodes and $\textrm{CFL} = 0.1$.}
    \label{fig:TGV}
\end{figure}

In this case, the problem of pressure oscillations does not have equally serious consequences, and as such all schemes were able to carry out the calculations up to the desired end time. Analyzing Fig.~\ref{fig:entropy_integral_TGV}, it is confirmed the result that the schemes based on the internal energy equation are able to reduce the spurious entropy production (see~\cite{Coppola_JCP_2019, DeMichele_C&F_2023}); comparing the original JP scheme to the modified JP-H, it appears that also the addition of the PEP property may provide a benefit, even though to a lesser extent.
Comparing the results of A$\rho$-H$e$, KEEP and KEEP-H, they all are of the same order of magnitude, with only minor deterioration in entropy conservation  for the KEEP-H scheme, in line with the theory reported in \cite{DeMichele_JCP_2023}, being mathematically less close to an exact EC scheme. 
Adding an additional term in the series expansions leads to a machine zero error in entropy conservation for both AEC${}^{(1)}$ and KEEP-H${}^{(1)}$.

An additional insight into the reliability of the schemes can be obtained through the study of the time evolution of thermodynamic fluctuations which are expected to stabilize around a constant value after an initial transient, as in the case of inviscid isotropic homogeneous turbulence~\cite{Honein_JCP_2004,Pirozzoli_JCP_2010,Coppola_JCP_2019}.
The evolution of temperature fluctuations $ \langle T '\rangle $ is displayed in Fig.~\ref{fig:fluc_T_TGV}.
The schemes based on the internal energy equation are able to keep the value of the fluctuations constant; the JP-H scheme seems to produce an improvement in the behavior of the JP method, but the fluctuations still present an increase with time. Similar behavior could also be seen through the analysis of density fluctuations $ \langle \rho '\rangle $ (not shown).
\vspace{1cm}

\section{Conclusions}
In this study, we have discussed the characteristics of a spatial discretization of the compressible Euler equations that are necessary for it to possess the PEP property.
General conditions have been given for the discretization of the momentum equation and of the energy equation, be it pressure, internal or total energy; considering these requirements, schemes from the literature have been analyzed.
The classical JP scheme~\cite{Pirozzoli_JCP_2010} and the KEEP scheme \cite{Kuya_JCP_2018} have been modified to obtain the PEP property by using the harmonic mean for the interpolation of density in the numerical fluxes, which was found to be the correct counterpart to the use of the arithmetic mean on internal energy.

The numerical test of a two-dimensional density wave has confirmed the capability of the newly introduced schemes to maintain unchanged the initially constant fields of pressure and velocity, as desired.
Moreover, the numerical error on entropy conservation has also been analyzed, including the test of the three-dimensional Taylor-Green vortex, to make sure that the modifications did not corrupt the effectiveness of the original schemes.
The KEEP-H scheme showed an error of the same order of magnitude as the one of the original KEEP.
On the other hand, the JP-H scheme showed an improvement when compared to the JP scheme on which it was based.
Another positive change for this scheme was observed by considering thermodynamic fluctuations. However, the introduction of the PEP property was not enough to make them stabilize around a constant value.

These findings serve as a demonstration of the techniques that can be employed to instill the PEP property in numerical schemes that may not already exhibit it, showcasing the feasibility of making such modifications with minimal changes in current compressible flow solvers.

\section*{Acknowledgments}
We acknowledge the CINECA award under the ISCRA initiative, for the availability of high-performance computing resources and support.
\vspace{1cm}

\bibliographystyle{model1-num-names}
\bibliography{Biblio_KEP_Compr}

\end{document}